\documentclass[twocolumn,prl,twocolumn,superscriptaddress, nofootinbib]{revtex4-1}

\usepackage[T1]{fontenc}
\usepackage[latin9]{inputenc}
\usepackage{xcolor}
\usepackage{pdfcolmk}
\usepackage{mathrsfs}
\usepackage{amsmath}
\usepackage{amssymb}
\usepackage{graphicx}
\usepackage{wasysym}
\PassOptionsToPackage{normalem}{ulem}
\usepackage{ulem}
\usepackage[unicode=true,pdfusetitle,
 bookmarks=true,bookmarksnumbered=false,bookmarksopen=false,
 breaklinks=false,pdfborder={0 0 0},pdfborderstyle={},backref=false,colorlinks=true]
 {hyperref}
\hypersetup{
 citecolor=blue,urlcolor=blue}

\makeatletter

\providecolor{lyxadded}{rgb}{0,0,1}
\providecolor{lyxdeleted}{rgb}{1,0,0}

\DeclareRobustCommand{\lyxsout}[1]{\ifx\\#1\else\sout{#1}\fi}

\usepackage{bbm}
\usepackage[final]{changes}
\usepackage{ulem}

\makeatother

\begin{document}
\title{Dressed Majorana fermion in a hybrid nanowire}
\author{Guo-Jian Qiao }
\affiliation{Beijing Computational Science Research Center, Beijing 100193, China}
\author{Xin Yue}
\affiliation{Beijing Computational Science Research Center, Beijing 100193, China}
\author{C. P. Sun}
\email{suncp@gscaep.ac.cn}

\affiliation{Beijing Computational Science Research Center, Beijing 100193, China}
\affiliation{Graduate School of China Academy of Engineering Physics, Beijing 100193,
China}
\affiliation{School of Physics, Peking University, Beijing 100871, China}
\begin{abstract}
The low-energy theory of hybrid nanowire systems fails to define Majorana
fermion (MF) in the strong tunneling and magnetic field strength.
To address this limitation, we propose a holistic approach to define
MF in which the quasi-excitation in nanowire and superconductor constitutes
together its own ``antiparticles''. This definition is general, beyond
the constraint presented in the low-energy theory. It reveals that
the Majorana phase depends not only on the chemical potential and
Zeeman energy in nanowire but also on those of superconductor, and
that the mismatch of chemical potential leads not to observe MF. Such
a broader perspective provides more specific experimental guidance
under various conditions.
\end{abstract}
\maketitle
\textit{Introduction}.---The observation of Majorana fermion (MF)
in condensed matter systems is a prerequisite for realizing topological
quantum computation \citep{A_Yu_Kitaev_2001,Sau_2010_Generic_New_Platform,Alicea_2012,Lutchyn2018,Prada2020}.
As one of the promising platforms, the semiconductor nanowire with
appreciable spin-orbit coupling coupled to an \textit{s}-wave superconductor
(SC), as considered to display MF \citep{Lutchyn_2010,Oreg_2010,Mourik_2012},
mainly based on a low-energy effective theory \citep{Alicea_2012,Stanescu_2011_MF_in_semiconductor_nanowires,Stanescu_2017_Proximity-induced_low-energy_renormalization}.
Such a low-energy effective theory predicts that the topological phase
of MF can be implemented in experiments by only adjusting the chemical
potential and applied magnetic field in the nanowire, directly independent
on the proximity superconducting materials.

The above low-energy effective theory is derived by eliminating the
virtual process of exchange of electron in nanowire between Bogoliubov
quasi-particles in SC, and a constant \textit{s}-wave pairing is induced
in the nanowire. But the validity of the low-energy theory is limited
due to the following three approximations \citep{Stanescu_2011_MF_in_semiconductor_nanowires,Stanescu_2017_Proximity-induced_low-energy_renormalization,supplementalmaterials}: 

(i) the \textit{s}-wave SC is assumed at a periodic boundary to ignore
the boundary effect, which indicates that the edge state of MF is
only localized at two ends of the nanowire; 

(ii) the energy-independent coupling spectral between the nanowire
and SC in wide-band limit results in the induced constant pairing
term; 

(iii) only low-energy effects of superconducting self-energy correction
are considered, which limits the effective theory to the weak tunneling
strength between the nanowire and SC.

What is more, it has also been pointed out that the effective Kitaev
model cannot be reduced when the Zeeman splitting and tunneling strength
are comparable to the superconducting gap due to the failure of perturbation
method \citep{Qiao_2022}, which leaves unresolved the question whether
MF can be defined in the strong magnetic field and coupling region.

\begin{figure}
\includegraphics[scale=0.6]{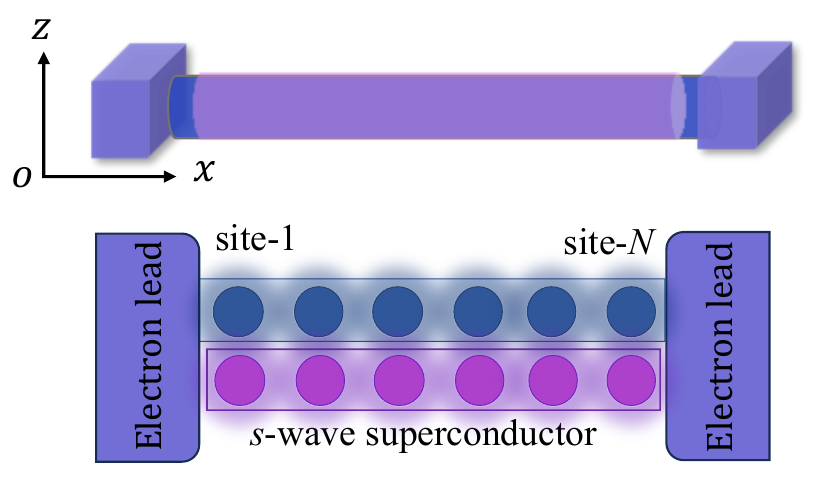}

\caption{The semiconductor nanowire is partially or fully covered by the superconducting
shell, which can be characterized by the one-dimensional lattice model
of nanowire coupled to an \textit{s}-wave superconductor. And the
spectrum of its differential conductance is obtained by connecting
electron leads at the ends of the nanowire.}
\label{hybrid-system}
\end{figure}

However, in the mainstream hybrid system used to observe MF experimentally
such as InSb/InAs nanowire partially or fully covered by aluminum
film, the coupling strength between nanowire and SC has reached the
intermediate or even strong coupling schemes under the strong magnetic
field \citep{Das2012_Zero_bias_peaks,Marco_2021_Nontopological_zero-bias,Zhang2018,Vaiti_2020,Lutchyn2018}.
Besides, some artificially experimental signature of MF fitting a
phase diagram given by the above low-energy theory have caused great
controversy and were even retracted so that the validity of the low-energy
theory should be doubted \citep{Zhang2018,Gazibegovic2017,Vaiti_2020,He_2017}.
Therefore, it is necessary to address the following issues:

(i) how to generally define MF in a hybrid system at any magnetic
field and tunneling strength ? 

(ii) what is universal Majorana topological phase like? Does it also
depend on the relevant parameters of SC ?

(iii) under what conditions, the topological phase of MF obtained
by the low-energy effective Hamiltonian is valid ?

In this paper, by a minimal hybrid model \citep{Lutchyn2018} that
a one-dimensional semiconductor nanowire couples to a one-dimensional
\textit{s}-wave SC {[}Fig. \ref{hybrid-system}{]}, we propose a holistic
perspective to define the dressed Majorana fermion (DMF), namely,
DMF is composed of the electron and hole excitations in nanowire dressed
by quasi-excitations in SC. Here, we treat the excitations in nanowire
and SC equally, rather than directly eliminating quasi-excitation
in SC, as has been done in the low-energy effective theory. Therefore,
such a definition of DMF holds for any tunneling and magnetic field
strength. 

The edge state of DMF in open boundary is localized not only at two
ends of nanowire but also of SC, and its distribution in nanowire
and SC is determined by the tunneling strength. As a signature of
DMF, we show that the $2e^{2}/h$ zero-bias peak in the differential
conductance spectrum is also obtained by connecting electron leads
to nanowire.

We demonstrate that the universal Majorana phase depends not only
on the chemical potential and Zeeman energy in nanowire but also in
SC. The low-energy phase diagram given by the effective theory {[}\textit{it
depends only on chemical potential and magnetic field in nanowire}{]}
is only an approximate result of the universal Majorana phase in the
weak tunneling scheme.

Lastly, the universal Majorana phase is applied to the experimental
hybrid InSb/InAs-Al system, and it is suggested that the mismatch
of chemical potentials between nanowire and SC hinders the observation
of MF.

\textit{Theoretical Model}.---The semiconductor nanowire with the
Rashba spin-orbit coupling $\alpha_{w}$ is described by $N$-sites
lattice model as \citep{Stanescu_2011_MF_in_semiconductor_nanowires,Stoudenmire_2011,Jelena_2016}
\begin{align}
H_{w} & =\sum_{n=1}^{N}d_{n}^{\dagger}\,[(t_{w}-\mu_{w})\sigma_{0}+h_{w}\sigma_{x}]\,d_{n}\nonumber \\
 & -\sum_{n=1}^{N-1}d_{n}^{\dagger}\,(\frac{t_{w}}{2}\sigma_{0}+\frac{i\alpha_{w}}{2}\sigma_{y})\,d_{n+1}+\mathrm{H.c.},\label{eq:nanowire-H}
\end{align}
where $d_{n}:=[d_{n\uparrow},d_{n\downarrow}]^{T}$ is a annihilation
operator of the electron in site-$n$ with spin $\uparrow,\downarrow$,
and $\sigma_{0,x,y}$ are the $2\times2$ identity and Pauli matrices.
Here, $\mu_{w}$ is the chemical potential, $t_{w}$ is the hopping
strength between nearest sites in the nanowire, and $h_{w}$ is the
Zeeman splitting caused by an external magnetic aligned with the nanowire
axis.

The semiconductor nanowire is in contact with an \textit{s}-wave SC
or wrapped by superconducting shell (such as Al \citep{Das2012_Zero_bias_peaks,Zhang2018}
and Pb \citep{Cao_2022,Jiang_2022_PbTe_Pb} shell), and such a SC
is also characterized by the one-dimensional lattice model: \citep{Jelena_2017_Finite_size_effects}
\begin{align}
H_{s} & =\sum_{n=1,\sigma}^{N}(t_{s}-\mu_{s})c_{n,\sigma}^{\dagger}c_{n,\sigma}-\frac{t_{s}}{2}[c_{n,\sigma}^{\dagger}c_{n+1,\sigma}+\mathrm{H.c.}]\nonumber \\
 & +\sum_{n=1}^{N}[\Delta_{s}c_{n,\uparrow}^{\dagger}c_{n,\downarrow}^{\dagger}+h_{s}c_{n\uparrow}^{\dagger}c_{n\downarrow}+\mathrm{H.c.]}.\label{eq:sc-H}
\end{align}
Similarly, $\mu_{s}$ is the chemical potential, $t_{s}$ is hopping
strength, $h_{s}$ is the Zeeman energy \citep{Clogston_1962,Maki1964,Reeg_2017,Marcus_Effective_g_2018}
and $\Delta_{s}$ is a pairing strength for the \textit{s}-wave SC. 

And the interaction between nanowire and\textit{ }SC is via single
electron tunneling: $H_{t}=-\sum_{n,m,\sigma}T_{nm}[d_{n\sigma}c_{m\sigma}^{\dagger}+c_{m\sigma}d_{n\sigma}^{\dagger}],$
with the tunneling strength $T_{nm}$ between the site-$n$ of the
nanowire and site-$m$ in SC. Below, we consider that the tunneling
strength for the same sites $n=m$ is equal but ignored between different
sites $n\neq m$, i.e., $T_{nm}\equiv\delta_{mn}T$. 

Notice that the Hamiltonian of the hybrid nanowire can be diagonalized
as $H=H_{w}+H_{s}+H_{t}=1/2\sum_{E}E\,\eta_{E}^{\dagger}\eta_{E}$,
and the corresponding quasi-particle operator is
\begin{equation}
\eta_{E}=(\mathbf{u}_{E}^{w})^{\dagger}\cdot\mathbf{d}+(\mathbf{v}_{E}^{w})^{\dagger}\cdot(\mathbf{d}^{\dagger})^{T}+(\mathbf{u}_{E}^{s})^{\dagger}\cdot\mathbf{c}+(\mathbf{v}_{E}^{s})^{\dagger}\cdot(\mathbf{c}^{\dagger})^{T},\label{eq:quasi-particle}
\end{equation}
where $\mathbf{d}:=[d_{1},\ldots,d_{N}]^{T}$ and $\mathbf{c}:=[c_{1},\ldots,c_{N}]^{T}$
with $c_{n}:=[c_{n\uparrow},c_{n\downarrow}]^{T}$. And the eigen-wave
function is $\Psi_{E}=[\mathbf{u}_{E},\mathbf{v}_{E}]^{T}$ with $\mathbf{u}_{E}=[\mathbf{u}_{E}^{w},\mathbf{u}_{E}^{s}]^{T}$
and $\mathbf{v}_{E}=[\mathbf{v}_{E}^{w},\mathbf{v}_{E}^{s}]^{T}$,
where $\mathbf{o}_{E}^{\alpha}=[\mathrm{o}_{E,1}^{\alpha},\ldots,\mathrm{o}_{E,N}^{\alpha}]^{T}$
with $\mathrm{o}_{E,n}^{\alpha}=[\mathrm{o}_{E,n\uparrow}^{\alpha},\mathrm{o}_{E,n\downarrow}^{\alpha}]^{T}$
respctively represent the wave function of electrons $\mathrm{o}\equiv\mathrm{u}$
and holes $\mathrm{o}\equiv\mathrm{v}$ in the nanowire and SC for
$\alpha=w,s$. 

\textit{Dressed MF}.---We define the dressed Majorana fermion as
\begin{equation}
\eta_{E}=\eta_{E}^{\dagger}\Longrightarrow\mathbf{u}_{E}^{w}=(\mathbf{v}_{E}^{w})^{*},\;\mathbf{u}_{E}^{s}=(\mathbf{v}_{E}^{s})^{*},
\end{equation}
where the quasi-excitation constitutes its own ``antiparticles''.
Further, the particle-hole symmetry requires $(\mathbf{u}_{-E}^{w})^{*}=\mathbf{v}_{E}^{w}$
and $(\mathbf{u}_{-E}^{s})^{*}=\mathbf{v}_{E}^{s}$. Thus it is proved
that only the zero-energy quasi-particle is Majorana fermion \citep{supplementalmaterials}.
Then, we need to give the conditions for the existence of such a zero-energy
wave function.

For the zero-enrgy wave function $E=0$ ($\mathbf{u}=\mathbf{v}^{*}$,
where subscript ``$0$'' is omitted), we decompose $\mathbf{u}$ into
real and imaginary parts $\mathbf{u}=\mathbf{u}^{(r)}+i\mathbf{u}^{(i)}$.
It is easy to verify that the electron wave function's real and imaginary
parts in nanowire satisfy
\begin{equation}
[(\mathbf{H}_{s}+\lambda\mathbf{P}_{s})\cdot\mathbf{H}_{w}-\mathbf{T}^{2}]\cdot\mathbf{u}_{w}^{(\lambda)}=0,\label{eq:real-part-eq}
\end{equation}
and the electron wave function in SC is $\mathbf{u}_{s}^{(\lambda)}=-\mathbf{T}^{-1}\cdot\mathbf{H}_{w}\cdot\mathbf{u}_{w}^{(\lambda)}$.
Here, $\lambda=\pm1$ corresponds to the real and imaginary parts
of $\mathbf{u}_{\alpha}^{(\lambda)}$, that is $\mathbf{u}_{\alpha}^{(1)}\equiv\mathbf{u}_{\alpha}^{(r)}$
and $\mathbf{u}_{\alpha}^{(-1)}\equiv\mathbf{u}_{\alpha}^{(i)}$ with
$\alpha=w,s$. And the $2N\times2N$ Hamiltonian matrices of the nanowire
$\mathbf{H}_{w}$, tunneling interaction $\mathbf{T}$ and SC ($\mathbf{H}_{s}$
and $\mathbf{P}_{s}$) are given in \textit{supplemental materials}
\citep{supplementalmaterials}.

We assume that $\mathrm{u}_{w,n}^{(r)}\equiv\xi^{n}[a_{\uparrow},a_{\downarrow}]^{T},1\leq n\leq N$,
where $\xi$ is a complex number, and $a_{\uparrow},a_{\downarrow}$
are undetermined coefficients. By substituting $\mathrm{u}_{w,n}^{(r)}$
into Eq. (\ref{eq:real-part-eq}), it is proved that $\xi$ is root
of $8$-order equation of one variable $f_{\lambda}(\xi)=0$ with
$\lambda=1$, and $a_{\uparrow},a_{\downarrow}$ are determined for
an given $\xi$ \citep{supplementalmaterials}. Thus such a electron
wave function in nanowire becomes $\mathrm{u}_{w,n}^{(r)}=\sum_{j=1}^{8}\alpha_{j}\xi_{j}^{n}[a_{j\uparrow},a_{j\downarrow}]^{T},$
where $\alpha_{j}$ are arbitrary superposition coefficients. 

The boundary and normalization conditions of the zero-energy wave
function $\mathbf{u}^{(r)}$ require that $f_{1}(\xi)=0$ must have
three roots less than $1$ as the total site number approaches infinity
$N\rightarrow\infty$, namely
\begin{equation}
f_{1}(\xi_{j})=0,\exists\:\xi_{j},j=1,2,3,\:|\xi_{j}|<1\label{eq:condition}
\end{equation}
At this time, $\mathbf{u}_{w}^{(r)}$ is uniquely determined, and
the wave function in SC is obtained by $\mathbf{u}_{s}^{(r)}=-\mathbf{T}^{-1}\cdot\mathbf{H}_{w}\cdot\mathbf{u}_{w}^{(r)}$. 

Similarly, the imagine part of the wave function $\mathbf{u}_{w}^{(i)}$
is obtained as $\mathrm{u}_{w,n}^{(i)}=\sum_{j=1}^{3}\beta_{j}\xi_{j}^{-N+n+1}[b_{j\uparrow},b_{j\downarrow}]^{T}$
for the given $\beta_{j}$ and $b_{j\sigma}$ with $\sigma=\pm1$,
and $\mathbf{u}_{s}^{(i)}=-\mathbf{T}^{-1}\cdot\mathbf{H}_{w}\cdot\mathbf{u}_{w}^{(i)}$.
Here, we have utilized the function relation $f_{\lambda=1}(\xi)=f_{\lambda=-1}(1/\xi)=0$
with $\xi\neq0$ \citep{supplementalmaterials}.

Known $\mathbf{u}^{(r)}$ and $\mathbf{u}^{(i)}$, DMF is rewritten
as
\begin{equation}
\begin{aligned}\gamma_{1} & =\sum_{n\sigma}\mathrm{u}_{w,n\sigma}^{(r)}(d_{n\sigma}+d_{n\sigma}^{\dagger})+\mathrm{u}_{s,n\sigma}^{(r)}(c_{n\sigma}+c_{n\sigma}^{\dagger}),\\
\gamma_{N} & =\sum_{n\sigma}\mathrm{u}_{w,n\sigma}^{(i)}(i[d_{n\sigma}^{\dagger}-d_{n\sigma}])+\mathrm{u}_{s,n\sigma}^{(i)}(i[c_{n\sigma}^{\dagger}-c_{n\sigma}]).
\end{aligned}
\label{eq:M-operator}
\end{equation}
It is seen from Eqs. (\ref{eq:condition}, \ref{eq:M-operator}) that
$\gamma_{1}$ and $\gamma_{N}$ correspond to the localized edge state
at site-$1$ and site-$N$ of the nanowire and SC. Such a MF is a
compsite of electron and hole exication in nanowire dressed by the
quasi-exication of SC {[}the last two terms of (\ref{eq:M-operator}){]},
and it is different from MF defined by the low-energy effective theory,
where the quasi-excitation in SC has been eliminated. 

Apparently, there is another case where the real and imaginary part
of zero-energy wave function are respetively localized near site-$N$
and site-$1$ of the nanowire and SC {[}$|\xi_{j}|>1$ in Eq. (\ref{eq:condition}){]}.
In short, a zero-energy wave function exists only when the function
$f_{1}(\xi)$ has three complex roots, and their magnitudes are greater
than or less than $1$. Thus, the parameter range in the presence
of the edge state of DMF is
\begin{equation}
\prod_{k}[(h_{w}+Z_{k}h_{s})^{2}-(\epsilon_{k}^{w}-\mu_{w}-Z_{k}(\epsilon_{k}^{s}-\mu_{s}))^{2}-(Z_{k}\Delta_{s})^{2}]<0,\label{eq:Mzm-phase}
\end{equation}
where $\epsilon_{k}^{s}:=t_{\alpha}(1-\cos k)$, and $k$ only takes
$0$ and $\pi$. The correction factor $Z_{k}:=T^{2}/(E_{k,+}^{s}E_{k,-}^{s})$
embodies the dressed effect of SC for the nanowire, where the excitation
energy of quasi-particle in SC is included as $E_{k,\pm}^{s}=\sqrt{[\epsilon_{k}^{s}-\mu_{s}]^{2}+\Delta_{s}^{2}}\pm h_{s}$
\citep{Alicea_2012}. Such a edge state of DMF depends not only on
the chemical potential $\mu_{w}$ and Zeeman energy $h_{w}$ of the
nanowire, but also on the chemical potential $\mu_{s}$ and Zeeman
energy $h_{s}$ in SC. And it can be proved that the conditions for
the existence of the Majorana edge state (\ref{eq:Mzm-phase}) is
consistent with the topological phase of DMF (Majorana phase) given
by the topological invariant $\text{sgn}(\mathrm{Pf}[\mathscr{H}_{\gamma}(0)])\text{sgn}(\mathrm{Pf}[\mathscr{H}_{\gamma}(\pi)])<0$
\citep{supplementalmaterials}. 

\begin{figure}
\includegraphics[scale=0.55]{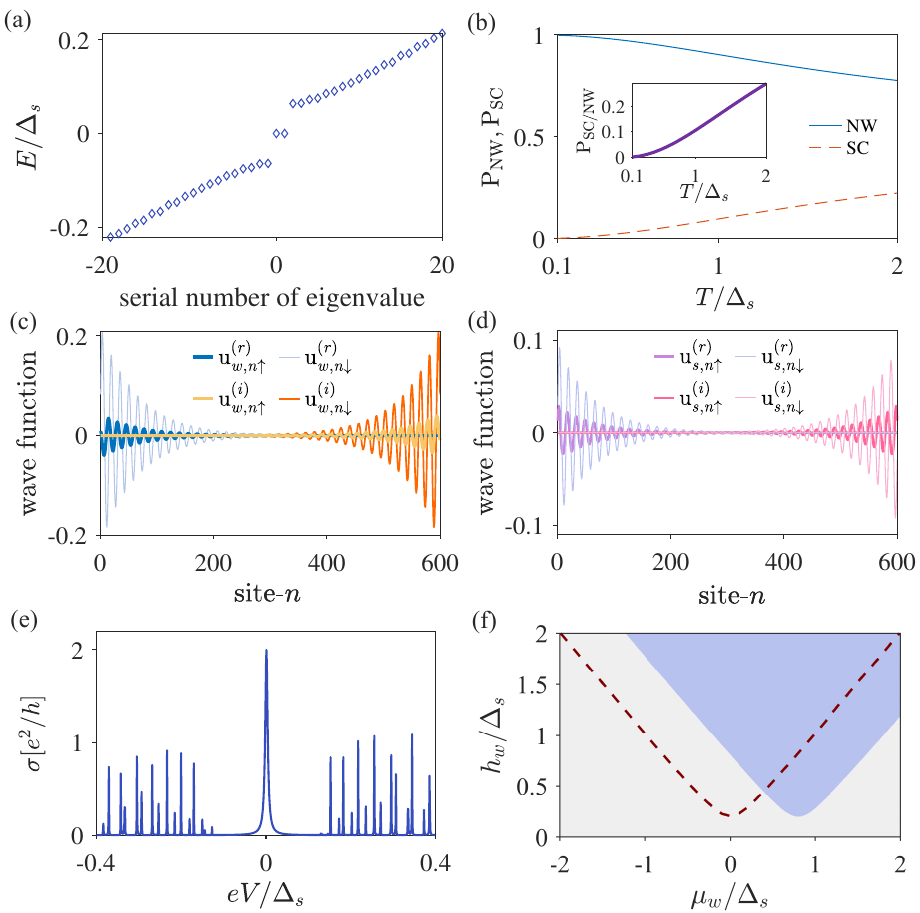}

\caption{(a) The zero-energy and excitation-energy modes in the hybrid system.
(b) The zero-energy wave function $\mathrm{P}_{\mathrm{SC}},\mathrm{P}_{\mathrm{NW}}$
in nanowire and SC, and their proportion $\mathrm{P}_{\mathrm{SC}}/\mathrm{P}_{\mathrm{NW}}$
change as the tunneling strength enhances. (c, d) The distribution
of the localized zero-mode wave functions at each site-$n$ in nanowire
and SC. (e) The zero bias peak in differential spectrum by connecting
electron leads at the ends of the nanowire. The parameters are set
as $N=600,t_{w}=12\Delta_{s},t_{s}=10\Delta_{s},\mu_{w}=0,\mu_{s}=4\Delta_{s},h_{w}=1.5\Delta_{s},h_{s}=0,\alpha_{w}=T=1.5\Delta_{s}.$
(f) The general Majorana phase (light purple region) deviates from
the phase diagram determined by the low-energy theory (red dashed
line) in $\mu_{w}-h_{w}$ space with the correction factor $Z_{0}=0.2$
and $h_{s}=0.1h_{w}$.}
\label{Fig:DMF-phase}
\end{figure}

In this Majorana phase, the energy spectrum of hybrid system indeed
displays the zero-energy modes of DMF, and there is a gap between
the zero and excitation energy {[}Fig. \ref{Fig:DMF-phase}(a){]}.
The edge state of DMF is not only localized at two ends in the nanowire,
but also in SC {[}Fig. \ref{Fig:DMF-phase}(c, d){]}. And the proportion
of zero-energy wave function distributed in the nanowire and SC $\mathrm{P}_{\mathrm{SC}}/\mathrm{P}_{\mathrm{NW}}$
increases with the enhance of tunneling strength $T$ , where the
zero-energy wave function of the nanowire and SC are measured by $\mathrm{P}_{\mathrm{NW}}=\sum_{n\sigma}|\mathrm{u}_{w,n\sigma}|^{2}+|\mathrm{v}_{w,n\sigma}|^{2}$
and $\mathrm{P}_{\mathrm{SC}}=\sum_{n\sigma}|\mathrm{u}_{s,n\sigma}|^{2}+|\mathrm{v}_{s,n\sigma}|^{2}$
with normalization condition $\mathrm{P}_{\mathrm{SC}}+\mathrm{P}_{\mathrm{NW}}=1$
{[}Fig. \ref{Fig:DMF-phase}(b){]}. This shows that the more zero-energy
wave function penetrates SC from the nanowire as the tunneling strength
increases. However, the low-energy effective theory has ignored the
zero-energy wave function in SC since the superconductor is considered
as a periodic boundary. In the intermediate, especially strong tunneling
schemes $T>\Delta_{s}$, the zero-energy wave function in SC is comparable
to that in nanowire {[}$T\sim2\Delta_{s},\:\mathrm{P}_{\mathrm{SC}}/\mathrm{P}_{\mathrm{NW}}\sim30\%${]},
and the boundary effect of SC cannot be ignored. 

It is known that MF determined by the low-energy theory will result
in the zero-bias peak (ZBP) with the height $2e^{2}/h$ at zero temperature
\citep{Qiao_2022,Li_2014,Flensberg_2010}. For DMF in the hybrid nanowire,
the $2e^{2}/h$ ZBP as a signature of DMF also appears by connecting
electron leads at the ends of the nanowire {[}see Fig. (\ref{hybrid-system})
and Fig. \ref{Fig:DMF-phase}(e){]} \citep{supplementalmaterials}.

Typically, the bandwidth of the nanowire and SC as the maximum energy
scale is much larger than their Zeeman energy, chemical potential
and the superconducting gap $t_{\alpha}\gg\mu_{\alpha},h_{\alpha},\Delta_{s}$
with $\alpha=w,s$. Then the universal Majorana phase (\ref{eq:Mzm-phase})
becomes
\begin{equation}
(h_{w}+Z_{0}h_{s})>\sqrt{(\mu_{w}-Z_{0}\mu_{s})^{2}+(Z_{0}\Delta_{s})^{2}}.\label{eq:appro-tp}
\end{equation}
When the dressed effect of SC for the nanowire (the correction factor
$Z_{0}$) is small so that $Z_{0}h_{s},Z_{0}\mu_{s}\ll1$, the Majorana
phase returns to the low-energy phase diagram $h_{w}>\sqrt{\mu_{w}^{2}+\Delta^{2}}$
with $\Delta:=Z_{0}\Delta_{s}$ \citep{Alicea_2012,Lutchyn2018}.
In this regard, the correction factor can be regraded as the ratio
of the induced pairing strength $\Delta$ to the superconducting gap
$Z_{0}=\Delta/\Delta_{s}$, which characterizes the different tunneling
schemes. By increasing the tunneling strength between nanowire and
SC, the correction factor gradually achieves a stronger tunneling
scheme $Z_{0}\rightarrow1$ so that the Zeeman energy $\tilde{h}_{w}:=h_{w}+Z_{0}h_{s}$
and chemical potential $\tilde{\mu}_{w}:=\mu_{w}-Z_{0}\mu_{s}$ in
the nanowire is significantly modified. This results in the Majorana
phase deviating significantly from the previous low-energy phase diagram,
especially for the more considerable superconducting chemical potentials
and Zeeman energy {[}Fig. \ref{Fig:DMF-phase}(f){]}. Therefore, the
Majorana phase (\ref{eq:appro-tp}) is more general, the fragile low-energy
topological phase is just an approximate result of which in weak tunneling
scheme $Z_{0}\ll1$. 

It is worth noting that the low-energy phase diagram\textit{ depends
only on chemical potential $\mu_{w}$ and magnetic field $h_{w}$
in nanowire}, which is considered as the solid theoretical basis to
observe MF in experiments so far. However, the universal Majorana
phase also depends significantly on the chemical potential $\mu_{s}$
and magnetic field $h_{s}$ in SC {[}see Eq. (\ref{eq:appro-tp}){]},
and it does have an impact on the observation of MF in an actual hybrid
nanowire system.

\textit{Dependence of Majorana phase on the chemical potential and
magnetic field in SC}---For the hybrid InSb/InAs-Al system, the superconducting
gap of aluminum film is $\Delta_{s}=0.34\,\mathrm{meV}$, and the
Zeeman splitting in SC is an order of magnitude smaller than that
in the nanowire due to the difference in Land\'{e} factor $h_{s}\sim0.05-0.15h_{w}$
\citep{Cao_2022,Reeg_2017,Marcus_Effective_g_2018,Lutchyn2018}. From
the induced energy gap observed in experiments $\Delta=0.2\,\mathrm{meV}$
\citep{Zhang2018}, the correction factor can be estimated as $Z_{0}=\Delta/\Delta_{s}\simeq0.6$.
Therefore the magnetic field required for the Majorana phase by (\ref{eq:appro-tp})
is obtained as
\begin{equation}
B>\frac{2\sqrt{(\mu_{w}-0.6\mu_{s})^{2}+(0.6\Delta_{s})^{2}}}{\mu_{\mathrm{B}}(\mathrm{g}_{w}+0.6\mathrm{g}_{s})}.
\end{equation}
with the Bohr magneton $\mu_{\mathrm{B}}$ and the Land\'{e} factor
$\mathrm{g}_{\alpha},\alpha=w,s$ of nanowire and SC. Since the critical
magnetic field of the aluminum film is $2\,\mathrm{T}$ \citep{Cao_2022},
the maximum difference in chemical potential between the nanowire
and SC cannot be greater than $|\mu_{w}-0.6\mu_{s}|<2.4/0.9\,\mathrm{meV}$
for the InSb\textbackslash InAs nanowire, where the Land\'{e} factors
are taken as $\mathrm{g}_{\mathrm{InSb}}=40,\mathrm{g}_{\mathrm{InAs}}=15$
and $\mathrm{g}_{s}=2$ \citep{Cao_2022,Lutchyn2018}. For example,
if the superconducting chemical potential is $1\mathrm{eV}$ \citep{Karsten_2018},
the chemical potential of the InSb\textbackslash InAs nanowire must
be adjusted to $597/599\lesssim\mu_{w}\apprle602/600\:\mathrm{meV}$.
Only within such a small range of the chemical potential, MF is possibly
observed. The smaller the external magnetic field, the smaller the
above window of chemical potential of nanowire for the Majorana phase.
Therefore, the mismatch of the chemical potentials in nanowire and
SC will prevent the Majorana topological phase. This may understand
why the signature of zero-bias peak in the current hybrid nanowire
systems is non-topological \citep{Marco_2021_Nontopological_zero-bias,Zhang2018,Zhang_2022}.

Based on the above facts, there are some remarks for observing MF
in experiments. (i) The magnetic field should be modest strength.
An excessively large magnetic field will inhibit the superconducting
gap \citep{Liu_2017_Andreev_bound_states,mohapatra2019observation,Pan_2020,Pan_2024},
while a smaller magnetic field will require higher control accuracy
of the chemical potential in nanowire and SC. (ii) The mismatch of
chemical potential highlighted above requires that the chemical potential
in the nanowire is adjusted in an extensive range to achieve the Majorana
phase, which may produce the current between the nanowire and SC due
to the drastic change of their potential difference. This is not conducive
to observe MF. (iii) The instability of (especially the larger) superconducting
chemical potential as other experimental parameters are adjusted also
destroys the signature of MF. Therefore, it is almost impossible to
observe the prominent MF signature robust to the large-range external
magnetic field and the gate voltage in the huge mismatch of chemical
potential.

\textit{Conclusion.---}We define the dressed Majorana fermion (DMF)
in any tunneling strength and magnetic field for the hybrid nanowire.
Under the open boundary, the edge state of DMF is localized at both
ends of the nanowire and superconductor (SC). And we obtain the general
topological phase of DMF, which is determined by the magnetic field
and the chemical potential in the nanowire and SC. We clarify the
validity of the low-energy phase diagram. Namely, it is only an approximate
result of the DMF phase in the weak tunneling scheme. We also point
out that in the mainstream hybrid InSb/As-Al system, the mismatch
in chemical potential between nanowire and SC makes MF challenging
to observe \citep{Das2012_Zero_bias_peaks,Zhang2018,Lutchyn2018,Marco_2021_Nontopological_zero-bias}.

Our theory provides a new method to define MF in the hybrid system
analytically. Namely, we treat the quasi-excitation in the nanowire
and superconductor equally instead of treating SC as an environment
to provide the proximity effect for the nanowire. Therefore, such
a definition of DMF is general and applicable to any parameters of
any chosen nanowire and SC materials, including the strong tunneling
strength and magnetic field.

Note that although \textit{s}-wave SC is described by a one-dimensional
lattice model in the above discussion, such a method of defining Majorana
fermion can be generalized to more hybrid systems, such as the two-dimensional
multi-band nanowire coupled to superconducting shell \citep{Lutchyn_2011_in_Multiband_Nanowires,Stanescu_2011_MF_in_semiconductor_nanowires,Jelena_2017_Finite_size_effects,Cole_2016},
and two-dimensional topological insulator-superconductor system \citep{Fu_2008_SC_Insulator,Wang_2015_SC_Insultor,Yue_2023}. 

\vspace{0.6em}

The authors appreciate quite much for the helpful discussion with
Sheng-Wen Li in BIT, and Dong E. Liu, Zhan Cao and Gu Zhang in BAQIS.
This study is supported by the National Natural Science Foundation
of China (NSFC) (Grant No. 12088101) and NSAF (Grant No. U2230402).

\bibliographystyle{apsrev4-2}
\bibliography{Refs}

\end{document}